\documentclass[10pt]{article}

\usepackage[a4paper,margin=0.68in]{geometry}
\usepackage[T1]{fontenc}
\usepackage[utf8]{inputenc}
\usepackage{lmodern}
\usepackage{microtype}
\usepackage{booktabs}
\usepackage{array}
\usepackage{enumitem}
\usepackage{url}
\usepackage{hyperref}
\usepackage{ragged2e}
\usepackage{xcolor}
\usepackage{dblfloatfix}
\usepackage{amsmath}

\hypersetup{
  colorlinks=true,
  linkcolor=black,
  citecolor=black,
  urlcolor=black
}

\setlist[itemize]{leftmargin=*, itemsep=1pt, topsep=2pt}
\setlist[enumerate]{leftmargin=*, itemsep=1pt, topsep=2pt}
\newcommand{\claim}{\mathsf{claim}}
\newcommand{\event}{\mathsf{event}}

\title{\textbf{An Evidence Model for Agentic Processes: Evidence Claims, Trust Assumptions, and Policy Assessment}}
\author{Arslan Br\"omme\\
\textit{CISSP, CISM, CISA, CAISE}\\
Independent Researcher\\
\texttt{arslan.broemme@aviomatik.de}}
\date{Draft v0.9.0.6 -- 8 September 2026}

\begin{document}
\twocolumn[
\maketitle
\vspace{-1.2em}
\begin{center}
\fbox{%
\begin{minipage}{0.94\textwidth}
\centering
\textit{Preprint / working paper. This version is a work in progress and may be updated. Comments are welcome.}
\end{minipage}}
\end{center}
\vspace{0.5em}
]

\begin{abstract}
Agentic AI systems increasingly exchange messages, invoke tools, request approvals, hold structured decision sessions, and modify shared artifacts. Logs and anchors can make selected records tamper-evident, but they can also mislead if their evidentiary meaning is implicit: a hash does not establish semantic truth, a signature does not establish authorization, and an external anchor does not establish capture completeness. This paper proposes an evidence claim model for agentic processes. It distinguishes artifact integrity, temporal existence, provenance, approval evidence, declared ordering, capture claim, relevance claim, deliberation traceability, monitoring claim, anchoring authorization claim, policy assessment claim, risk treatment claim, mitigation implementation claim, and management response claim. Semantic validity is treated as a recurring limitation. The model maps these claims to mechanisms, assumptions, limitations, and threats, and situates them in an agent organization with functional CEO agent, executive, operational, evidence, and audit roles, plus a plan--do--check--act-inspired management response loop. The contribution is conceptual: it does not validate a particular implementation, prevent all failures, or automate legal compliance. It provides a vocabulary for stating which claims an agentic black box can support, which claims it cannot establish, and which controls are required around it.
\end{abstract}
\vspace{0.7em}

\section{Introduction}

Autonomous and semi-autonomous AI agents increasingly operate as participants in organizational processes and may also be arranged as agent organizations with internal management, specialist, operational, evidence, and audit roles. They exchange instructions, draft recommendations, invoke tools, update artifacts, and request or receive human approvals. They may also participate in structured discussions in which specialized agents compare recommendations, risks, policy constraints, and dissenting views before an action is executed. These activities create an auditability problem: after an incident, compliance review, or management escalation, an agent organization, its human operators, or the responsible legal entity may need to reconstruct not only what was stored, but also which evidentiary claims can legitimately be made about the stored records.

The author's earlier architecture paper proposed a product- and vendor-neutral black-box architecture for agentic processes, based on the pipeline capture, canonicalize, hash, anchor, certify, and verify~\cite{broemme2026blackbox}. That paper introduced the architectural pattern and anchoring pipeline. The present paper substantially extends it by specifying the evidentiary semantics and limits of the resulting records, including capture-completeness, decision-session, audit, risk-treatment, mitigation-implementation, and management-response claims. The earlier architecture is treated as prior work and motivation, not as independent external validation.

The motivating question is deliberately narrow: \emph{which evidentiary claims can an agentic black box support?} The answer is not binary. A record can be integrity-verifiable but semantically false. It can be signed by an agent identity but still violate policy. It can be anchored externally and yet omit a suppressed event. For that reason, agentic evidence systems need explicit evidence claims and explicit trust assumptions.

The novelty is not a new hash construction, ledger protocol, or agent runtime, but a claim-oriented decomposition of agentic evidence with explicit limitations and trust assumptions. The paper makes three contributions: an evidence claim structure for agentic process records, a mapping from mechanisms such as hashes, signatures, sequence references, historian classifications, audit-agent findings, risk and mitigation records, and anchors to supported evidentiary properties, and an agent-organization model that segregates operational execution, evidence production, audit assessment, and management response.

This position and architecture paper does not present an empirical performance, security, or compliance evaluation. It proposes a conceptual model and uses a simplified governance example to illustrate evidentiary claims and their limits.

\section{Related Work}

The model builds on long-standing work on digital time-stamping, secure audit logs, canonicalization, transparency logs, remote attestation, and threshold signatures. Haber and Stornetta introduced cryptographic time-stamping for digital documents~\cite{haber1991timestamping}. RFC 3161 defines a time-stamp protocol that can provide evidence that a datum existed before a particular time~\cite{rfc3161}. Secure audit logging addresses tamper evidence and post-compromise reconstruction for logs on untrusted machines~\cite{schneier1999secureaudit}. JSON canonicalization is relevant because evidence records must be represented in a repeatable form before hashing or signing. The JSON Canonicalization Scheme is an informational example of such a method, not the only possible approach~\cite{rfc8785}.

Transparency-log systems such as Certificate Transparency demonstrate externally auditable append-only logs and Merkle-tree inclusion evidence in a certificate-specific context~\cite{rfc9162}. Remote attestation provides a vocabulary for evidence, claims, verifiers, and relying parties about the state of a computing environment~\cite{rfc9334}. Threshold signatures such as FROST can distribute trust for sensitive signing operations, but their value depends on key generation, participant independence, threshold assumptions, and operational governance~\cite{rfc9591}. NIST policy states that federal agencies may use SHA-2 family hash functions, including SHA-256 and SHA-512, for applications that employ secure hash algorithms~\cite{nistHashPolicy}. These mechanisms support different assurance claims. None of them, by itself, establishes semantic truth, capture completeness, or organizational compliance.

The threat model is also related to work on indirect prompt injection, where malicious instructions embedded in external content can influence LLM-integrated applications and agent behavior~\cite{greshake2023indirectprompt}. AI governance and cybersecurity regulation provide context for risk management, record-keeping, human oversight, incident handling, and management accountability. For high-risk AI systems within the relevant scope, the EU AI Act contains provisions on risk management (Art. 9), record-keeping (Art. 12), human oversight (Art. 14), and deployer obligations (Art. 26), but these provisions do not apply to every agentic system. For entities within its scope, NIS2 covers management governance (Art. 20), cybersecurity risk-management measures (Art. 21), and reporting obligations (Art. 23). NIS2 does not establish a general evidentiary model for AI agents. Applicability depends on system classification, intended purpose, operator role, entity, sector, service, transitional rules, and national transposition~\cite{euai2024,nis22022}.

\section{System and Governance Model}

The model assumes a deliberately simple agent organization: a CEO agent at the top, executive agents beneath the CEO agent, and employee or operational agents beneath them. The terms CEO agent and executive agent denote functional roles inside the illustrative agent organization, not legal persons, corporate organs, or independent legal authorities. The CEO agent receives audit reports, reviews risk-treatment proposals, and initiates process improvements as an internal management and evidence role. In real deployments, binding decisions, legal accountability, and final risk acceptance remain with the responsible human officers or legal entity.

Agentic workflows are owned by an operational executive agent, such as a CTO agent, CIO agent, or Head-of-AI-Operations agent. The evidence function is placed under a separate executive agent responsible for security, risk, compliance, or AI governance. The audit function is either an internal audit-agent function or a separate assurance function with reporting access to the CEO agent and, where required, to human oversight.

This separation matters because an evidence system should not be controlled only by the same agents and runtime that it is supposed to observe. If operational agents, logging, evidence classification, anchoring keys, audit assessment, and management reporting are all inside the same trust domain, the resulting evidentiary value is weaker. In smaller teams, open-source projects, or decentralized multi-organization settings, the same separation can be approximated through independent maintainers, separate keys, external audit services, read-only observers, or technically isolated evidence components rather than formal agent-executive roles.

The following roles are used throughout the paper.

\begin{itemize}
  \item \textbf{Operational agents} execute tasks, communicate with users or other agents, invoke tools, and update artifacts. Each operational agent has evidence duties, for example to emit events for approvals, privileged tool calls, and policy exceptions.
  \item \textbf{Agentic decision session} is a structured deliberation among specialized agents and, where required, human approvers. It produces an agentic decision record containing topic, participants, policy context, arguments, dissenting views, uncertainty, and outcome.
  \item \textbf{Logger} captures the event stream and stores raw or normalized records in an append-only local evidence store.
  \item \textbf{Historian} classifies selected events as decision-relevant, policy-relevant, incident-relevant, or audit-relevant. The historian is not a truth oracle. Instead, it signs relevance claims. Its classification output is itself recorded as an evidence record.
  \item \textbf{Evidence supervisor} monitors whether evidence capture policies are applied, checks for missing expected events, and escalates capture gaps.
  \item \textbf{Evidence controller} canonicalizes records, checks signatures and policy metadata, computes hashes or Merkle roots, and approves anchoring.
  \item \textbf{Anchoring service} publishes commitments to an external evidence layer, such as a public blockchain, timestamping service, transparency log, or permissioned ledger. The anchoring wallet is a governed technical component of this service, not a free-standing decision authority.
  \item \textbf{Audit-agent} evaluates evidence records against security policies, control requirements, and defined workflow rules.
  \item \textbf{CEO agent and executive agents} receive audit reports, prepare management responses, propose improvements, accept or reject risks within delegated authority, and update policies or processes. Binding legal authority remains with the responsible human officer or legal entity.
\end{itemize}

The organizational pattern can be summarized as follows:

\begin{center}
\begin{minipage}{0.94\linewidth}
\small
\begin{verbatim}
CEO agent
|- Executive AI Operations agent
|  `- Employee / operational agents
|     and decision sessions
|- Exec. Security / Risk / AI Governance agent
|  `- Logger, Historian, Evidence Supervisor,
|     Evidence Controller, Anchoring Service
`- Internal Audit / Assurance agent
   `- Audit-agent and reports to CEO agent
\end{verbatim}
\end{minipage}
\end{center}

The exact reporting lines may vary. The central requirement is that evidence production and audit assessment remain distinguishable from operational execution, both inside the agent organization and in the surrounding human governance structure.

\section{Threat Model}

The threat model distinguishes three groups: internal threats, external threats, and threats against the evidence infrastructure itself.

\subsection{Internal Threats}

Internal threats originate from deployed agents, tools, or organizational users inside the agentic system. They include agents that intentionally or unintentionally violate policies, suppress events, manipulate communication records, bypass logging mechanisms, or execute unauthorized tool calls. Such behavior may result from malicious objectives, misconfiguration, hallucination, over-optimization, defective tool integration, or inadequate human oversight.

Internal threats are especially relevant because the agent whose action is being evaluated may also be the component asked to report the action. Therefore, self-reporting by operational agents is useful, but insufficient. It creates a self-reporting claim, not a completeness guarantee.

\subsection{External Threats}

External threats include malicious users, external agents, compromised systems, adversarial documents, manipulated tool outputs, identity spoofing, replay attempts, and prompt injections~\cite{greshake2023indirectprompt}. In these cases, a malicious external input may influence an internal agent to follow unauthorized instructions, ignore a policy, misuse a tool, or rely on corrupted context.

For evidence purposes, prompt injection is not only a prevention problem. It is also a reconstruction problem. In the proposed model, an organization may later need to show which external content was accessed, which instructions were embedded in that content, which agent processed it, which policy checks were applied, and which actions followed.

\subsection{Threats Against Evidence Infrastructure}

A third threat group targets the evidence infrastructure itself. Examples include bypassing the logger, causing the historian to misclassify events, preventing the evidence supervisor from detecting missing records, compromising the evidence controller, misusing anchoring keys, delaying anchoring, or selectively anchoring only benign records.

A related concern is the evaluative component itself. The audit-agent occupies a privileged position: it interprets policies and issues findings that management relies on, yet it is itself an agent subject to misconfiguration, compromise, or manipulation of its inputs. An audit-agent finding is therefore a claim, not a ground truth, and its own evidentiary chain requires the same scrutiny as the records it evaluates.

The model does not eliminate these threats. It requires them to be made explicit. Its value depends on segregation of duties, key governance, authenticated capture, monitoring, append-only buffers, independent verification, organizational escalation, and human review of selected high-impact findings. In particular, capture completeness can only be assessed relative to an expectation model. Expected-event definitions, mandatory capture policies, sequence counters, tool-gateway logs, independent observers, and remote attestation can improve confidence, but they cannot establish that no off-system event occurred.

A simple operational indicator for mandatory-event coverage can be expressed as a matching rate. Let $M$ be the number of expected mandatory events with at least one validated observation, and let $N_{\mathrm{expected}}$ be the number of mandatory events expected under an independently defined and versioned evidence-capture policy, workflow specification, or tool-gateway rule. Then
\[
C = \frac{M}{N_{\mathrm{expected}}}.
\]
Here, $C$ denotes mandatory-event coverage. It is not a completeness proof. It is the quantitative signal underlying the evidence supervisor's monitoring claim in Table~\ref{tab:claims}. $C=1$ means that every expected mandatory record has a validated observation, while $C<1$ indicates a capture gap. Multiple validated observations of the same expected event count once toward $M$, with consistency assessed separately. The critical value remains $N_{\mathrm{expected}}$: it must not be inferred only from the observed evidence stream, because missing records would then disappear from the denominator. Without such an expectation model, capture coverage cannot be meaningfully computed. The ratio is not evidence of global completeness. Its usefulness depends on the independence and quality of the expectation model.

\section{Evidence Claim Model}

An evidence claim is a statement about what a retained record, artifact, process step, or audit result can support. A claim is not merely a fact stored in a log. It is a structured assertion whose strength depends on mechanism and assumptions. We model an evidence claim as:

\[
\begin{aligned}
\claim = (&\event, \text{property}, \text{mechanism}, \text{assumptions},\\
& \text{limitation}, \text{threat scope}).
\end{aligned}
\]

The \emph{event} may be an agent message, human approval, tool call, policy exception, decision session record, audit finding, risk treatment record, mitigation implementation record, or management decision. The \emph{property} is the evidentiary property being asserted. The \emph{mechanism} is the technical or organizational mechanism that supports the claim. The \emph{assumptions} state what must be true for the claim to hold. The \emph{limitation} states what the claim does not establish. The \emph{threat scope} states which threat the claim is meant to address. The tuple is a conceptual notation. It is not yet a full formal semantics for claim composition, conflicting claims, confidence levels, or support states such as supported, partially supported, and not supported.

For example, consider a human approval event. A canonicalized record, a SHA-512 digest, and an external anchor can support the claim that the retained approval record still matches the committed representation and that the commitment existed no later than the anchor time. They do not establish that the approval was legally valid, that the human understood all consequences, or that no other unrecorded discussion occurred.

\begin{table*}[!t]
\centering
\footnotesize
\renewcommand{\arraystretch}{1.04}
\begin{tabular}{p{0.17\textwidth} p{0.245\textwidth} p{0.245\textwidth} p{0.215\textwidth}}
\toprule
\textbf{Mechanism} & \textbf{Supported evidence claim} & \textbf{Trust assumption} & \textbf{Limitation} \\
\midrule
{\RaggedRight Canonicalization + hash\par} & Artifact integrity: retained artifact matches committed digest. & Canonicalization is deterministic and the retained artifact is available. & Does not establish truth, intent, correctness, or completeness. \\
External anchor & Temporal existence: commitment existed no later than inclusion or finality time. & Anchor system is independently verifiable and available. & Does not establish exact event creation time or causal order. \\
Agent or human signature & Provenance: record was signed by a specific key or identity. & Key binding, authentication, and key protection are reliable. & Does not establish signer intent, honesty, or legal responsibility. \\
Recorded approval & Approval evidence: a specific approval or rejection decision was recorded and linked to policy context. & Approver identity, role binding, authorization scope, validity interval, delegation state, and policy version are separately verifiable. & Does not by itself establish wisdom, legal sufficiency, or that all consequences were understood. \\
Sequence references & Declared ordering: records reference previous events or a sequence. & Sequence generation is controlled and gaps are detectable. & Does not establish real-world causality or absence of parallel events. \\
Logger capture & Capture claim: event passed through the logging path. & Logger is correctly integrated and cannot be silently bypassed. & Does not establish that all relevant events were captured. \\
Historian classification & Relevance claim: event was classified as policy-, decision-, audit-, or incident-relevant. & Historian identity, policy basis, and classification time are recorded. & Does not establish objective importance or semantic validity. \\
Decision session record & Deliberation traceability: participants, rationale, dissent, and outcome were recorded. & Required participants and decision rules are defined and enforced. & Does not create organizational authority by itself. \\
Evidence supervisor & Monitoring claim: capture policy was checked against observed event stream. & Expected events can be defined independently of agent self-report. & Unknown off-system events may remain invisible. \\
Evidence controller & Anchoring authorization claim: record was validated and released for anchoring. & Controller rules, roles, and keys are governed independently. & Does not establish content correctness or upstream capture completeness. \\
Audit-agent finding & Policy assessment claim: selected records were evaluated against policies. & Policies are machine-readable, current, and correctly interpreted. & Does not create compliance or establish legal sufficiency. \\
Risk treatment record & Risk treatment claim: risk, owner, treatment option, due date, residual risk, and management decision were documented. & Risk workflow, owner assignment, and management authority are defined. & Does not establish that the risk assessment was complete or objectively correct. \\
Mitigation implementation record & Mitigation implementation claim: a specific mitigation action was documented as implemented or partially implemented. & Implementation evidence is captured from a reliable source and linked to the correct risk and mitigation. & Does not establish mitigation effectiveness or residual-risk reduction. \\
Management response record & Management response claim: report receipt, decision, or risk acceptance was recorded. & Executive action was captured and bound to the report. & Does not establish that the chosen measure was adequate. \\
\bottomrule
\end{tabular}
\caption{Evidence mechanisms, claims, assumptions, and limitations.}
\label{tab:claims}
\end{table*}

Table~\ref{tab:claims} illustrates the central design principle: the evidentiary question is not whether a system is verifiable in general, but which property is supported, by which mechanism, under which assumptions, and against which threat.

\section{Evidence Capture and Policy Assessment}

A practical agentic evidence system needs an evidence capture policy. Otherwise, evidence selection becomes arbitrary and the historian becomes too powerful. The policy should define mandatory capture, risk-based capture, and historian-nominated capture.

Mandatory capture should cover human approvals, rejected approvals, privileged tool calls, external system access, policy exceptions, security-relevant decisions, changes to important artifacts, incident-related communication, audit findings, risk treatment records, mitigation implementation records, and responses by the CEO agent or executive agents. Risk-based capture can apply to unusual agent-agent communication, conflicting recommendations, repeated failed tool calls, high-impact decisions, deviations from standard workflows, or structured decision sessions. Historian-nominated capture can preserve contextual discussions and rationale that are not mandatory but may later be relevant.

A minimal event record should include an event identifier, event type, actor, related agent or human role, policy identifier and policy version, locally asserted event time, prior-record reference, artifact hash, relevance classifications, and signatures. For the illustrative approval event \texttt{ae-0042}, the record would bind \texttt{human\_approval}, \texttt{human-analyst-01}, \texttt{remediation-agent-02}, \texttt{AI-GOV-07} version \texttt{1.3}, a created-at timestamp, a SHA-512 artifact hash, a prior event hash, and the actor, historian, and evidence-controller signatures.

\begin{center}
\begin{minipage}{0.96\linewidth}
\scriptsize
\begin{verbatim}
record_id: ae-0042
type: human_approval
actor: human-analyst-01
subject: remediation-agent-02
policy: AI-GOV-07@v1.3
created_at: 2026-09-06T15:55:00Z
related_records: [decision-session-009, tool-call-017]
artifact_hash: sha512:<digest>
previous_record_hash: sha512:<previous-digest>
classifications: [decision-relevant]
signatures: [actor, historian, evidence-controller]
anchor_reference: batch-2026-09-06-15
\end{verbatim}
\end{minipage}
\end{center}

The \texttt{created\_at} field represents locally asserted event time, whereas \texttt{anchor\_reference} points to later external inclusion or finality evidence. The same envelope can represent approval records, decision-session records, audit findings, risk treatment records, mitigation implementation records, and management responses by varying \texttt{type} and adding type-specific payload fields. Interoperability therefore does not require one universal schema for all agentic evidence. It requires a small common envelope, stable identifiers, policy-version binding, canonicalization rules, and clearly defined extensions for different record classes.

Typical extensions remain compact. A \texttt{decision\_session} record may add participants, inputs considered, arguments, dissent, uncertainty, and outcome. An \texttt{audit\_finding} record may add checked records, policy version, result, severity, and finding rationale. A \texttt{risk\_treatment} record may add risk identifier, likelihood, impact, owner, treatment option, due date, and residual risk. A \texttt{mitigation\_implementation} record may add implementation evidence, affected agents or tools, changed policy or workflow, verification status, and follow-up audit reference.

Historian classifications should be captured as separate evidence records that state which event was classified, which policy or heuristic was applied, what rationale was given, when the classification occurred, and which historian identity signed it. This prevents the historian from becoming an invisible relevance gatekeeper. Negative or low-relevance classifications may not all be anchored individually, but sampling, monitoring, or batch commitments can make selective suppression harder to hide.

A practical implementation must balance assurance against cost, privacy, and operability. Immediate anchoring gives stronger temporal granularity but increases cost and metadata exposure. Merkle batching reduces cost and leakage but weakens immediacy. Broader capture improves reconstruction but may create privacy, retention, and deletion constraints that require deployment-specific legal and policy analysis. More signatures and review steps improve accountability but add operational overhead. These trade-offs should be policy decisions rather than hidden implementation side effects.

The audit-agent consumes evidence records and evaluates them against security policies and control objectives. Its output is not compliance itself, but a signed policy-assessment claim over selected records. For example, it may state that a privileged tool call was preceded by a recorded human approval under policy \texttt{AI-GOV-07}, or that a required escalation event is missing. The finding should bind the checked record set, input commitments, policy version, audit-agent model or prompt configuration, and output commitment. This allows later reviewers to distinguish conformance under the rule in force at event time from later policy changes or audit-agent configuration changes.

\section{Decision Sessions and the Management Cycle}

Agentic governance requires not only evidence of actions, but also evidence of deliberation. Before high-impact or policy-sensitive actions, an agent organization may convene agentic decision sessions in which specialized agents discuss a decision. A planning agent may propose an action, a risk agent may evaluate impact, a security agent may identify abuse paths, a compliance agent may assess policy requirements, and a human approver may accept, reject, or escalate the recommendation where human oversight is required. The session should produce a structured decision record that captures topic, participants, inputs considered, policy context, arguments, dissenting views, confidence levels, required approvals, and outcome.

An agentic decision session does not create organizational authority by itself. It creates a documented decision process inside the agent organization. Legal, managerial, or policy authority remains with the responsible human role, organizational function, or legal entity. Its evidentiary value lies in preserving how the decision was discussed, what warnings were raised, and which rationale was available when approval or escalation occurred.

Audit-agent findings should feed a risk and mitigation cycle rather than remain isolated reports. Where an audit finding identifies a policy deviation, missing approval, insufficient evidence capture, unclear agent-agent communication, or exposure to prompt-injected content, the agent organization should create a risk treatment record. Such a record links the finding to a risk description, affected policy, proposed mitigation, responsible owner, due date, residual risk, and management decision. The implementation of mitigation measures should itself become part of the evidence stream. A mitigation implementation record can document whether the measure was implemented, which policy or workflow changed, which agents or tools were affected, and which follow-up verification is required.

This management response loop is plan--do--check--act-inspired. In the Plan phase, the agent organization defines security policies, evidence capture policies, risk treatment rules, and control objectives. In the Do phase, operational agents execute workflows and mitigation measures while evidence duties are applied. Agentic decision-session records document structured deliberation before or during operational action. In the Check phase, the evidence supervisor and audit-agent assess whether required records exist, whether policies were followed, and whether mitigation implementation can be verified. In the Act phase, the CEO agent or executive agents update policies, workflows, agent communication rules, tool permissions, or propose residual-risk acceptance for human ratification where required. Audit findings, risk treatment records, mitigation implementation records, and follow-up verification records therefore support the Check and Act phases, while decision-session records primarily support the Do phase. The resulting management decisions and policy changes become new evidence records and define the baseline for the next cycle.

\begin{center}
\small
\fbox{\begin{minipage}{0.92\linewidth}
Evidence records $\rightarrow$ audit-agent finding $\rightarrow$ risk treatment $\rightarrow$ management decision $\rightarrow$ mitigation implementation $\rightarrow$ follow-up verification $\rightarrow$ policy improvement.
\end{minipage}}
\end{center}

\section{Example: Agent Organization and Decision Session}

Consider an agent organization coordinating a remediation workflow. A planning agent proposes a remediation step, a risk agent evaluates impact, a security agent identifies a prompt-injection risk, and a compliance agent checks whether policy \texttt{AI-GOV-07} requires human approval. Where human oversight is required, a human approver accepts, rejects, or escalates the proposed action. An execution agent invokes the tool, and a verification agent checks the result. The workflow also processes an external document containing adversarial instructions that attempt to cause a policy bypass.

Before execution, the agents enter an agentic decision session. The decision record preserves the proposed action, external input, policy version, risk concern, prompt-injection warning, approval requirement, dissenting views, and final recommendation. The historian classifies the record as decision-relevant and, because adversarial content was involved, incident-relevant. The evidence supervisor checks that mandatory capture rules were triggered for external content, human approval, privileged tool use, and the decision session. The evidence controller canonicalizes records, verifies signatures, computes hashes, and submits a Merkle root to the anchoring service, which publishes the commitment. The audit-agent later evaluates whether human approval was recorded before execution and whether the mitigation workflow was followed.

Instantiating the claim tuple above for approval event \texttt{ae-0042} yields the following bounded evidence claim:

\begin{center}
\begin{minipage}{0.96\linewidth}
\scriptsize
\begin{verbatim}
Event: approval event ae-0042
Property: evidence of a recorded approval
Mechanism: signed approval record + sequence reference
           + anchored hash
Assumptions: approver key valid, capture path active,
             role record and AI-GOV-07 v1.3 available
Limitation: does not establish actual authorization,
            wisdom, or legal sufficiency by itself
Threat scope: later denial or alteration of approval record
\end{verbatim}
\end{minipage}
\end{center}

A linked mitigation record can instantiate the same structure:

\begin{center}
\begin{minipage}{0.96\linewidth}
\scriptsize
\begin{verbatim}
Event: mitigation record mitigation-031
Property: mitigation implementation claim
Mechanism: owner-signed implementation record
           + linked risk treatment record
           + follow-up verification reference
Assumptions: owner binding valid,
             implementation evidence captured,
             affected agents/tools correctly linked
Limitation: does not establish mitigation effectiveness or
            residual-risk reduction by itself
Threat scope: denial or alteration of mitigation status
\end{verbatim}
\end{minipage}
\end{center}

The CEO agent receives the audit report and prepares a management response within the agent organization. If the audit-agent identifies that external content was processed too close to a privileged execution path, the management layer creates a risk treatment record. A mitigation may require external content to be isolated, summarized by a separate agent, and prevented from directly influencing privileged tool calls. Implementation is tracked through a mitigation implementation record. A later audit-agent run verifies whether the updated workflow was applied in subsequent cases. Where legal authority, binding organizational change, or formal risk acceptance is required, the response must be reviewed or ratified by the responsible human officer or legal entity. The resulting evidence is useful but bounded: it can show that records, findings, risks, mitigations, and management responses existed in a committed form. It cannot establish that no unobserved communication occurred, that the mitigation was effective in all cases, or that the organization was legally compliant.

\section{Limitations and Future Work}

The proposed model improves evidentiary precision, but it does not solve all assurance problems. It does not guarantee truthful agent behavior, complete capture, correct policy interpretation, legal compliance, effective mitigations, or adequate management decisions. Events that are never captured cannot be recovered by later hashing. A compromised capture component can still record a sanitized representation. A historian can misclassify relevance. A valid signature can belong to a compromised key.

More fundamentally, the model does not resolve who assesses the assessor. If the audit-agent's findings are generated, transmitted, and stored through the same infrastructure it is meant to evaluate, an unaddressed regress remains: a compromised or misconfigured audit-agent could produce a well-formed, signed, anchored finding that nonetheless misrepresents the underlying evidence. Mitigating this requires the audit function to sit in a distinct trust domain from the evidence controller and anchoring service. At minimum, audit-agent output should use separately governed signing keys, bind immutable policy versions, record input and output commitments, document the model and prompt configuration used for the assessment, follow an independent capture path, and be subject to periodic human sampling against the underlying records. Future work should examine whether second-order evidence about the audit process is a proportionate control, or whether it merely relocates the same trust problem one level up.

The model also depends on practical governance choices. Key management, segregation of duties, retention rules, privacy constraints, access control, monitoring quality, and the independence of the anchoring layer affect the evidentiary strength of the system. Capture completeness remains relative: it can be assessed against defined expectations and independent signals, but cannot establish the non-existence of every unobserved communication or side channel. Blockchain anchoring is one possible externalization mechanism, but timestamping services, transparency logs, and permissioned ledgers may also be appropriate depending on threat model and regulatory environment~\cite{rfc3161,rfc9162}.

Future work should formalize an evidence claim language, including claim composition, conflict handling, confidence levels, and support states. It should define interoperable event schemas, evaluate canonicalization and Merkle batching strategies, test capture completeness under adversarial workflows, integrate remote attestation for evidence controllers, and study how audit-agent findings can be reviewed by human and external auditors. Schema registries, evidence profiles, and conformance tests for canonicalization and claim validation should also be investigated. Another open question is how external auditors can independently verify selected evidence without exposing sensitive agent communications, risk records, mitigation details, or business data.

Future work should also evaluate the model in realistic multi-agent deployment environments, including chat-operated gateways, tool-executing agent runtimes, and managed multi-agent settings. Experiments should instrument communication, tool invocation, approvals, audit-agent findings, and management responses, test whether expectation models for mandatory events can be defined, assess whether the matching-based coverage indicator $C$ produces useful audit signals, and examine whether adversarial workflows, missing approvals, or privileged tool misuse can be reconstructed. The objective is model usability under realistic communication, execution, and governance conditions, not benchmarking a specific product.

\section{Conclusion}

Agent organizations need more than logs. They need a precise understanding of which claims their evidence can support. This paper proposed a compact evidence model for agentic processes that separates artifact integrity, temporal existence, provenance, approval evidence, declared ordering, capture claim, relevance claim, deliberation traceability, monitoring claim, anchoring authorization claim, policy assessment claim, risk treatment claim, mitigation implementation claim, and management response claim. Semantic validity remains a recurring limitation rather than something established by anchoring or signing alone.

The model's main conclusion is simple: the relevant question is not whether agentic evidence is verifiable in general, but which evidentiary property is supported, by which mechanism, under which trust assumptions, and against which threat. A well-designed agentic black box should therefore not only preserve records. It should make the evidentiary meaning of those records explicit and support a closed management cycle from evidence capture and audit findings to risk treatment, mitigation implementation, follow-up verification, and policy improvement.

\vspace{-0.3em}
\noindent\fbox{%
\begin{minipage}{0.96\linewidth}
\scriptsize
\textbf{Declaration on the Use of AI Tools.} AI language models, in particular OpenAI GPT-5.6 and Anthropic Claude Sonnet 5, were used as tools during the preparation of this paper for language drafting, critical review, source checking, discussion of examples, and LaTeX/PDF artifact generation. The author remains solely responsible for the content, conceptual decisions, source selection, and final version.
\end{minipage}}

\end{document}